\documentclass[lettersize,journal]{IEEEtran}
\usepackage{amsmath,amsfonts}
\usepackage{algorithmic}
\usepackage{algorithm}
\usepackage{booktabs}
\usepackage{array}
\usepackage[caption=false,font=normalsize,labelfont=sf,textfont=sf]{subfig}
\usepackage{textcomp}
\usepackage{stfloats}
\usepackage{url}
\usepackage{amssymb}
\usepackage{verbatim}
\usepackage{graphicx}
\usepackage{xcolor}
\usepackage[section]{placeins}
\usepackage{placeins}
\usepackage{cite}
\usepackage[colorlinks=true,
            linkcolor=blue,
            citecolor=blue,
            urlcolor=black]{hyperref}
\makeatletter
\renewcommand{\@cite}[2]{%
  \textcolor{blue}{[{#1\if@tempswa , #2\fi}]}%
}
\makeatother
\begin{document}
\bstctlcite{IEEEexample:BSTcontrol}
\title{Compound Climate Risk-Aware Electricity Dispatch: Trajectory-Based Modeling for Power System Resilience}

\author{Feiyu~Cai,~\IEEEmembership{Graduate~Student~Member,~IEEE}, 
Jing~Qiu,~\IEEEmembership{Senior~Member,~IEEE},
Chenxi~Zhang,~\IEEEmembership{Member, IEEE},
Yi~Yang,~\IEEEmembership{Member, IEEE},
Yuechuan~Tao,~\IEEEmembership{Member, IEEE},
Xin~Lu,~\IEEEmembership{Senior~Member,~IEEE}
}

\markboth{IEEE Transactions on Power Systems}%
{Shell \MakeLowercase{\textit{et al.}}: A Sample Article Using IEEEtran.cls for IEEE Journals}


\maketitle

\begin{abstract}
In recent decades, emerging compound hazards in a changing climate have increasingly threatened power system resilience and posed new challenges to electricity dispatch. Conventional Value-at-Risk (VaR)-based risk-aware dispatch provides a tractable means of managing stochastic net demand deviations caused by distribution feeder outages. However, when applied to temporally compounding events such as tropical cyclone (TC)--blackout--heatwave, conventional net demand risk formulations may assign similar risk levels to a brief outage and an outage with the same instantaneous net demand impact that persists into a subsequent hazard, despite their different requirements for sustained operational flexibility. Moreover, directly incorporating compound risk into electricity dispatch models faces a structural mismatch: the risk is defined by the joint distribution of feeder restoration and subsequent hazards over exponentially many joint states, whereas dispatch models require risk formulations that are tractable in the scheduling decisions. To address these issues, this paper proposes a compound climate risk-aware dispatch framework with a trajectory-based compound risk model, which defines the loss on operating deviations determined by entire feeder connection trajectories, parameterized by uncertain restoration times and coupled with subsequent hazards. Analytical moment-generating functions and equivalent exponential-cone reformulations provide a compact representation of the resulting risk constraints, preserving the relevant dependencies without explicit joint-state enumeration. By leveraging a physics-based spatiotemporal TC--blackout--heatwave hazard model, case studies demonstrate the effectiveness of the proposed method with real-world time-series power outage data from Puerto Rico, including data collected during Hurricane Fiona in 2022.
\end{abstract}


\section{Introduction}
\label{sectionI}
\IEEEPARstart{E}{XTREME} events such as tropical cyclones (TCs), heatwaves and floods are becoming increasingly severe and more frequent under a changing climate, leading to numerous widespread catastrophic blackouts globally~\cite{bennett2021extending,zheng2025strategies}. Despite growing efforts to enhance power system resilience, climate-induced outages remain a major threat to modern socio-economic activities~\cite{ruan2026long}. AusNet reported a marked increase in blackout duration on major event days, following several severe climate extremes since 2020 in Australia~\cite{AusNet2025RegulatoryProposal}.
Compared with transmission networks, distribution networks are generally more vulnerable to climate extremes due to their lower infrastructure resistance (extensive distribution feeders) and higher penetration of renewable energy (solar and wind), both of which are highly exposed to environmental conditions~\cite{xu2024resilience}. For instance, Hurricane Fiona in 2022 damaged more than half of Puerto Rico's nearly 1000 distribution feeders, resulting in approximately US\$2.5 billion in economic losses~\cite{DOE2022Fiona,NOAA2023}. 

Meanwhile, as one of the critical solutions to achieve the net-zero emissions goal for energy systems, large-scale integration of distributed energy resources is reshaping the traditional distribution networks into active distribution networks involving both electricity producers and consumers~\cite{jacob2024real,zhang2024two,cai2026emission}. However, these environment-sensitive renewable energy resources are highly susceptible to climate extremes. During a thunderstorm in 2025, extensive cumulonimbus cloud cover in Western Australia's Interconnected System sharply reduced rooftop solar generation, triggering a 49\% operational demand surge within 40 minutes~\cite{bomThunderstorms}. When coupled with existing outage risk in distribution networks, the integration of such distributed resources further amplifies operational uncertainty, complicating efforts to maintain the system energy balance at the transmission network level during climate extremes. Therefore, with increasing distributed renewable integration and escalating climate risks, it is critical to incorporate the risk arising from distribution networks with large-scale distributed energy resources into system-level electricity dispatch during climate extremes~\cite{jiang2026climate}.

Considerable research effort has been devoted to risk management for electricity dispatch optimizations. Compared to robust optimization (RO) focusing on the worst scenario~\cite{chen2018distributionally}, stochastic optimization (SO), including scenario-based methods~\cite{du2018scenario} and Value-at-Risk (VaR)~\cite{rockafellar2000optimization} methods, ensures electricity dispatch satisfies system constraints, either deterministically or with a specified probability within sampled scenarios~\cite{wang2016risk}. Authors in~\cite{zhang2023unit} employed Conditional Value-at-Risk (CVaR) to characterize wind power uncertainty and embedded it into a power-to-gas-enabled integrated energy system unit commitment (UC) model for risk-aware dispatch.~\cite{zeng2024efficient} introduced Entropic Value-at-Risk (EVaR) to risk-constrained optimal power flow under wind power uncertainty to improve computational efficiency. In particular, the authors in~\cite{xu2024entropic} employed EVaR to improve the computational tractability of risk modeling involving numerous discrete feeder-outage states under extreme climate events. Furthermore, the Risk-aware Electricity Dispatch Under Climate Extremes with Renewable integration (REDUCER) model was developed in~\cite{xu2025riskaware} by embedding a climate--energy model into an EVaR-constrained UC formulation, thereby enabling day-ahead dispatch under extensive distribution network outages and distributed renewable generation uncertainty.

However, evolving climate change and ongoing transformation of power systems are creating increasingly complex interactions between climate hazards and grid operations. Such interactions can give rise to compound climate hazards, where multiple hazards occur simultaneously or sequentially and jointly exacerbate system impacts~\cite{zscheischler2018future}. For instance, the probability of long-duration heatwaves following strong TCs may increase sharply under future climate conditions~\cite{matthews2019emerging}. During TC-induced power outages, the impact of heatwaves would dramatically increase, as air conditioning is critical in reducing the vulnerability to extreme heat~\cite{feng2022tropical,feng2025hurricane}. Therefore, the emerging TC--blackout--heatwave compound hazard introduces a temporal compounding dimension to grid operation. 

Existing risk-aware operation models generally apply CVaR or EVaR directly to hourly outage-induced net demand deviations. These formulations capture the magnitude and probability of each deviation, but not how long the underlying outage is expected to last or whether it overlaps with a subsequent hazard during temporally compounding events. Consequently, a brief outage and an outage with the same instantaneous net demand impact that persists into a subsequent hazard may receive similar risk treatment, despite requiring different levels of sustained operational flexibility requirements and commitment decisions. On the other hand, directly incorporating complex compound risk into power system operating models presents a structural mismatch between the joint probabilistic representation of feeder restoration and subsequent hazards and the explicit algebraic constraints required for scheduling. The combinatorial growth of joint feeder states further complicates risk evaluation, limiting the integration of these risks into UC and dispatch decisions.

To address these issues, this paper proposes a compound climate risk-aware dispatch framework that incorporates the joint impacts of uncertain feeder restoration and subsequent hazards into post-event electricity dispatch under temporally compounding events. A trajectory-based compound risk model, together with its analytical representation and equivalent exponential-cone reformulation, is derived to embed the resulting compound risks into electricity scheduling. The main contributions of this paper are summarized as follows.
\begin{enumerate}
\item{A trajectory-based compound risk model is developed for compound climate extremes. Instead of applying risk measures to hourly net-demand deviations, it defines the loss on operating deviations determined by entire feeder connection trajectories, parameterized by uncertain restoration times and coupled with post-event weather scenarios. The model thus captures outage persistence, its overlap with subsequent hazards, and cross-feeder dependence induced by shared weather scenarios.}
\item{An exact analytical formulation of the compound risk is derived via conditional moment-generating function factorization, reducing the $|\mathcal{M}|(T+2)^{|\mathcal{I}|}$ joint scenario--restoration combinations to $|\mathcal{M}||\mathcal{I}|(T+2)$ feeder-level terms. The resulting nested log-sum-exp structure is shown to admit an equivalent exponential-cone representation, yielding convex risk constraints whose cone arguments are affine in the scheduling decisions.}
\item{The proposed framework embeds these risk constraints into a post-event electricity dispatch model with large-scale distributed renewable integration and energy storage, constraining the risks of generator outputs, line flows, ramping, and storage state of charge (SOC). The SOC risk accounts for energy deviations accumulated along restoration trajectories, explicitly linking outage persistence to the sustained flexibility required in the resulting mixed-integer exponential-cone program.}
\end{enumerate}

The remainder of this paper is organized as follows.
Section~\ref{sectionII} presents an overview of the proposed framework.
Section~\ref{sectionIII} develops the trajectory-based compound risk model and derives its analytical expressions and equivalent exponential-cone reformulations.
Section~\ref{sectionIV} formulates the risk-aware electricity dispatch model with large-scale distributed renewable integration and energy storage.
Section~\ref{sectionV} evaluates the proposed framework through case studies.
Section~\ref{sectionVI} concludes the paper.

\section{Overview of the Proposed Framework}
\label{sectionII}
\begin{figure}[!htbp]
\centering
\includegraphics[width=\columnwidth]{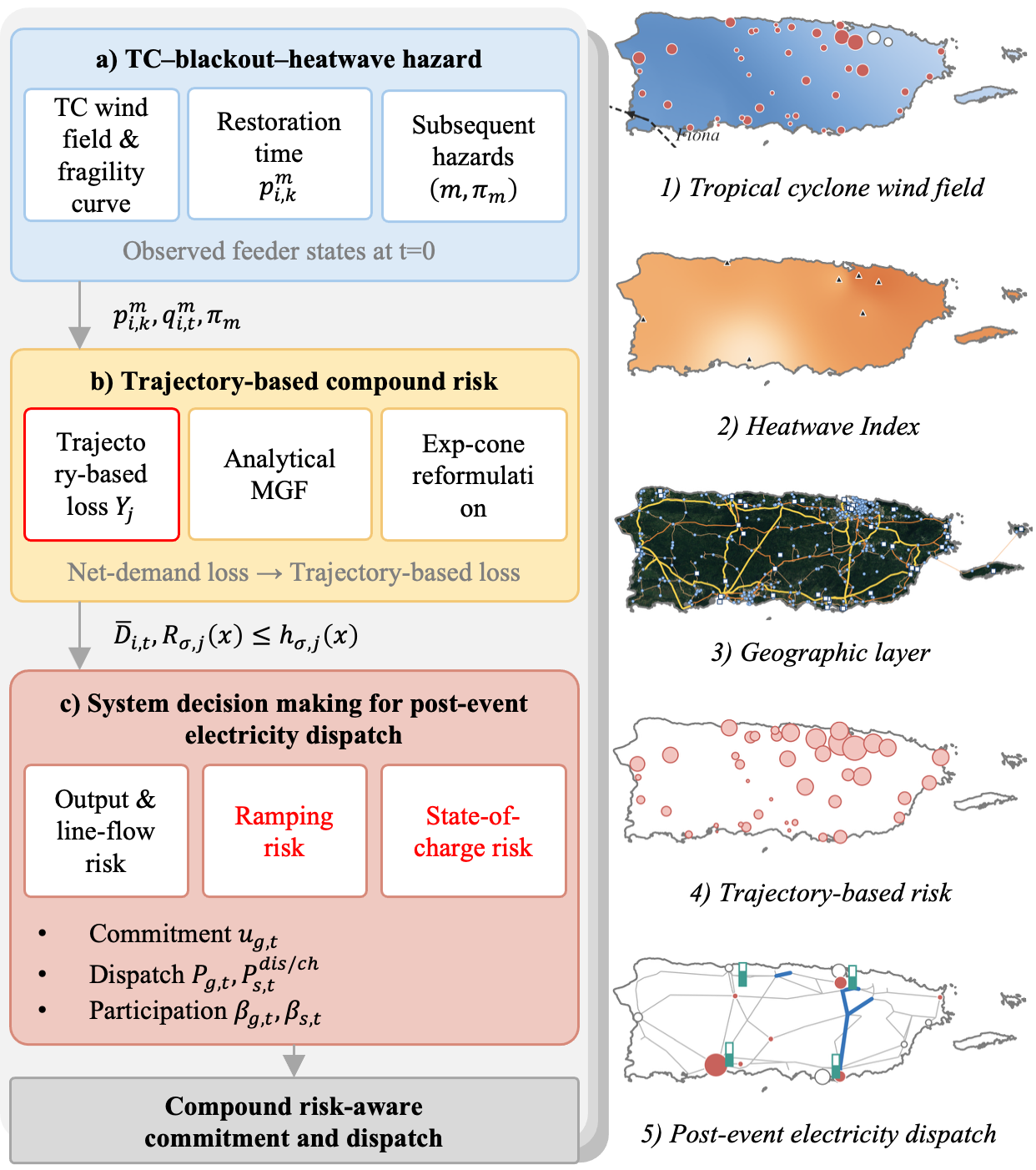}
\caption{Overview and data flow of the proposed framework.}
\label{fig.1}
\end{figure}

The proposed framework is illustrated in Fig.~\ref{fig.1}. It is organized as a sequential data flow through three modules, in which the outputs of each module serve as the inputs of the next. The scheduling decisions are made after the initial TC exposure, when the feeder connection states are observed, while the subsequent restoration times and hazards remain uncertain.
The compound hazard module (Fig.~\ref{fig.1}(a)) takes storm-track data, photovoltaic (PV) fragility functions, observed outage records, and post-TC temperature and humidity trajectories as inputs. The reconstructed wind field yields the PV damage probabilities $p_i^{\mathrm{PV}}$, the outage records provide the initial feeder states and calibrate the restoration-time probabilities $p_{i,k}^m$, and the meteorological trajectories define the subsequent hazard scenarios $m\in\mathcal M$ with probabilities $\pi_m$. Combining these with demand and irradiance attenuation models gives the scenario-dependent connected-state net demand $q_{i,t}^m$. The module thus passes $\{p_{i,k}^m, q_{i,t}^m, \pi_m\}$ to the risk modeling module.

The risk modeling module (Fig.~\ref{fig.1}(b)) uses these quantities to construct the trajectory-based loss. Each feeder's connection trajectory is determined by its restoration time and combined with $q_{i,t}^m$, yielding the expected connected net demand $\bar D_{i,t}$ and the trajectory-weighted contributions $B_{j,i,k}^m(x)$ and $c_j(x)$ for each operating deviation $Y_j$. The analytical moment-generating function and its exponential-cone reformulation then convert each EVaR term into conic constraints whose cone arguments are affine in the scheduling decisions. The module passes $\bar D_{i,t}$ and these risk constraints to the scheduling module.

The scheduling module (Fig.~\ref{fig.1}(c)) combines them with generator, storage, and network parameters in a post-event UC model. Here, $\bar D_{i,t}$ sets the scheduling baseline in the balance, reserve, and power-flow constraints, while the conic risk constraints limit the risks of generator outputs, line flows, ramping, and SOC. The outputs are the commitment schedules, generation and storage dispatch, and balancing participation factors.

\section{Trajectory-Based Compound Risk Modeling}
\label{sectionIII}
\subsection{Conventional EVaR Modeling of Net-Demand Deviations}
\label{sectionIII-A}
EVaR is a coherent risk measure defined through the moment-generating function (MGF) of a random loss~\cite{ahmadi2012entropic}. For a loss function $g(x,\omega)$, where $x$ denotes the decision vector and $\omega$ denotes an uncertainty realization, EVaR at confidence level $1-\alpha$, with $\alpha\in(0,1)$, is defined as:
\begin{equation}
\operatorname{EVaR}_{1-\alpha}(g(x, \omega))
=
\inf_{r>0}
\frac{1}{r}
\ln\left(
\frac{M_{g(x, \omega)} (r)}{\alpha}
\right)
\label{eq.1}
\end{equation}
where $M_{g(x, \omega)}(r)=\mathbb E[\exp(rg(x, \omega))]$ is the MGF of the loss function $g(x, \omega)$ evaluated at any $r>0$. 

For distribution feeder $i\in\mathcal I$ at time $t$, let $S_{i,t}\in\{0,1\}$ denote its operating state, where $0$ and $1$ indicate the failed and in-service states of the feeder, respectively, and let feeder failure probability $p_{i,t}^{\mathrm{out}}=\Pr(S_{i,t}=0)$. Let $q_{i,t}=P_{i,t}^{\mathrm{demand}}-P_{i,t}^{\mathrm{PV}}$ denote its net demand when in service, where $P_{i,t}^{\mathrm{demand}}$ and $P_{i,t}^{\mathrm{PV}}$ are the feeder load and distributed photovoltaic (PV) output, respectively. The actual net demand seen by the transmission system is $D_{i,t}=S_{i,t}q_{i,t}$, whose expectation $\mathbb E[D_{i,t}]=(1-p_{i,t}^{\mathrm{out}})q_{i,t}$. Its deviation from this expectation is:
\begin{equation}
\begin{aligned}
\Delta P_{i,t}^{\mathrm{net}}
&:=D_{i,t}-\mathbb E[D_{i,t}]\\
&=
\begin{cases}
-(1-p_{i,t}^{\mathrm{out}})q_{i,t},
& S_{i,t}=0,\\
p_{i,t}^{\mathrm{out}}q_{i,t},
& S_{i,t}=1
\end{cases}
\end{aligned}
\label{eq.2}
\end{equation}

Define the aggregate deviation as $\Delta P_t^{\mathrm{net}}=\sum_{i\in\mathcal I}\Delta P_{i,t}^{\mathrm{net}}$. Taking $\sigma\Delta P_t^{\mathrm{net}}$ as the loss in Eq.~\eqref{eq.1} yields the directional risk $\mathcal R_{\sigma,t}^{\mathrm{EV}}$. Under conditional independence of feeder states at each time $t$, the MGF factorizes across feeders, the conventional directional EVaR is given by:
\begin{equation}
\begin{aligned}
\mathcal R_{\sigma,t}^{\mathrm{EV}}
&:=
\operatorname{EVaR}_{1-\alpha}
\left(\sigma\Delta P_t^{\mathrm{net}}\right)\\
&=\min_{s_{\sigma,t}>0}
s_{\sigma,t}\sum_{i\in\mathcal I}
\ln\Bigg[
p_{i,t}^{\mathrm{out}}
\exp\left(
\frac{-\sigma(1-p_{i,t}^{\mathrm{out}})q_{i,t}}
{s_{\sigma,t}}
\right)\\
&+
(1-p_{i,t}^{\mathrm{out}})
\exp\left(
\frac{\sigma p_{i,t}^{\mathrm{out}}q_{i,t}}
{s_{\sigma,t}}
\right)
\Bigg]
-s_{\sigma,t}\ln\alpha
\label{eq.3}
\end{aligned}
\end{equation}
where $\sigma=+1$ and $\sigma=-1$ represent upward and downward net demand risks, respectively, and $s_{\sigma,t}$ is the reciprocal of the variable used in MGF.
\subsection{Trajectory-Based Compound Risk and Analytical Modeling}
\label{sectionIII-B}
The hourly formulation in Eq.~\eqref{eq.3} characterizes marginal net-demand risks but does not explicitly represent the joint net-demand trajectory induced by feeder restoration and subsequent hazards. In temporally compounding events, restoration determines which feeders remain disconnected when a subsequent hazard changes their connected-state net demands. We model these coupled uncertainties and derive an exact, compact risk expression for generator, line-flow, ramping, and SOC constraints without enumerating joint feeder states.

We consider an operating horizon $\mathcal T=\{1,\ldots,T\}$ following the initial hazard, with observed feeder states but uncertain restoration times and subsequent hazard evolution. Let $\tau_i$ denote the first period in which feeder $i$ is restored, with $\tau_i=0$ for initially in-service feeders and $\tau_i=T+1$ for feeders remaining disconnected throughout the horizon. Assuming that initial in-service and restored feeders remain connected throughout the remainder of the horizon, the operating state of feeder $i$ is given by:
\begin{equation}
\widetilde S_{i,t}=
\begin{cases}
0, & t<\tau_i,\\
1, & t\geq\tau_i
\end{cases}
\label{eq.4}
\end{equation}

Subsequent hazard evolution is represented by scenarios $m\in\mathcal M$ with probabilities $\pi_m$. Each scenario specifies the connected-state net demand $q_{i,t}^{m}=P_{i,t}^{\mathrm{demand},m}-P_{i,t}^{\mathrm{PV},m}$ and the conditional restoration probabilities $p_{i,k}^{m}=\Pr(\tau_i=k\mid m)$, $k\in\{0,\ldots,T+1\}$. All probabilities are conditional on the information available at the beginning of the horizon. Restoration times are assumed independent across feeders conditional on $m$, while each feeder's states across time are linked by the same $\tau_i$.

Under scenario $m$, the actual connected net demand is:
\begin{equation}
\widetilde D_{i,t}=\widetilde S_{i,t}q_{i,t}^{m}
\label{eq.5}
\end{equation}
The same restoration time determines the feeder's entire connection trajectory, while $q_{i,t}^{m}$ captures subsequent hazard effects. Taking the expectation over both uncertainties gives:
\begin{equation}
\bar D_{i,t}
:=
\mathbb E[\widetilde D_{i,t}]
=
\sum_{m\in\mathcal M}
\pi_m q_{i,t}^{m}
\sum_{k=0}^{t}p_{i,k}^{m}
\label{eq.6}
\end{equation}
Accordingly, the centered deviation in Eq.~\eqref{eq.2} becomes $\Delta\widetilde P_{i,t}^{\mathrm{net}}=\widetilde D_{i,t}-\bar D_{i,t}$,
with aggregate deviation $\Delta\widetilde P_t^{\mathrm{net}}=\sum_{i\in\mathcal I}\Delta\widetilde P_{i,t}^{\mathrm{net}}$.

To evaluate how compound net-demand uncertainty affects system operation, let $Y_j(x,\omega)$ denote the deviation of operating quantity $j$ from its scheduled value. Here, $j$ identifies a particular generator output, line flow, or intertemporal output change. This deviation can be expressed as:
\begin{equation}
Y_j(x,\omega)
=
\sum_{i\in\mathcal I}\sum_{t\in\mathcal T}
b_{j,i,t}(x)\Delta\widetilde P_{i,t}^{\mathrm{net}}(\omega)
\label{eq.7}
\end{equation}
where $x$ is the scheduling decision vector and $b_{j,i,t}(x)$ specifies how feeder $i$'s net-demand deviation at time $t$ contributes to operating deviation $Y_j$. These coefficients are determined by the operating equations. For a single-period output or line-flow deviation, only coefficients associated with that period are nonzero; for an SOC deviation, coefficients from all preceding periods enter the expression. Thus, Eq.~\eqref{eq.7} provides a unified representation of the random operating deviations whose risks are evaluated below. 

To derive an analytical formulation of compound operating risk, we first express the operating deviation in terms of feeder restoration times. Under hazard scenario $m$, define the weighted contribution of feeder $i$ restored in period $k$ and the mean correction as:
\begin{subequations}\label{eq.8}
\begin{align}
B_{j,i,k}^{m}(x)
&:=
\sum_{\substack{t\in\mathcal T\\t\geq k}}
b_{j,i,t}(x)q_{i,t}^{m}
\label{eq.8a}\\
c_j(x)
&:=
\sum_{i\in\mathcal I}\sum_{t\in\mathcal T}
b_{j,i,t}(x)\bar D_{i,t}
\label{eq.8b}
\end{align}
\end{subequations}
where $B_{j,i,k}^{m}(x)$ accounts for the feeder's weighted contribution after restoration, while $c_j(x)$ removes the contribution associated with the expected net demand. An empty sum is zero, giving $B_{j,i,T+1}^{m}(x)=0$. Combining Eq.~\eqref{eq.4}--\eqref{eq.8}, the operating deviation under scenario $m$ becomes:
\begin{equation}
\begin{aligned}
Y_j(x,\omega)
&=
\sum_{i\in\mathcal I}\sum_{t\in\mathcal T}
b_{j,i,t}(x)
\left(\widetilde S_{i,t}q_{i,t}^{m}-\bar D_{i,t}\right)\\
&=
\sum_{i\in\mathcal I}
\sum_{\substack{t\in\mathcal T\\t\geq\tau_i}}
b_{j,i,t}(x)q_{i,t}^{m}
-c_j(x)\\
&=
\sum_{i\in\mathcal I}B_{j,i,\tau_i}^{m}(x)-c_j(x)
\end{aligned}
\label{eq.9}
\end{equation}
Thus, the contribution of an entire feeder trajectory is determined by a single restoration time and the scenario-dependent net demand.

Let $\mathcal K=\{0,\ldots,T+1\}$ and $\boldsymbol\tau=(\tau_i)_{i\in\mathcal I}$. A vector $\boldsymbol k=(k_i)_{i\in\mathcal I}\in\mathcal K^{|\mathcal I|}$ specifies a joint restoration realization. For direction $\sigma\in\{-1,+1\}$ and $r>0$, the law of total expectation gives the following joint-state representation of the MGF:
\begin{equation}
\begin{aligned}
M_{\sigma Y_j}(r)
&=
\mathbb E\!\left[e^{r\sigma Y_j(x,\omega)}\right]\\
&=
\sum_{m\in\mathcal M}\pi_m
\sum_{\boldsymbol k\in\mathcal K^{|\mathcal I|}}
\Pr(\boldsymbol\tau=\boldsymbol k\mid m)\\
&\quad\times
\exp\!\left[
r\sigma\left(
\sum_{i\in\mathcal I}B_{j,i,k_i}^{m}(x)-c_j(x)
\right)
\right]
\end{aligned}
\label{eq.10}
\end{equation}
Direct evaluation of this expression would enumerate up to $|\mathcal M|(T+2)^{|\mathcal I|}$ joint scenario--restoration combinations. To eliminate this enumeration, we use the conditional independence of restoration times:
\begin{equation}
\Pr(\boldsymbol\tau=\boldsymbol k\mid m)
=
\prod_{i\in\mathcal I}p_{i,k_i}^{m}
\label{eq.11}
\end{equation}
together with the additive representation in Eq.~\eqref{eq.9}. The joint sum then factorizes as:
\begin{equation}
\begin{aligned}
M_{\sigma Y_j}(r)
&=
e^{-r\sigma c_j(x)}
\sum_{m\in\mathcal M}\pi_m
\sum_{\boldsymbol k\in\mathcal K^{|\mathcal I|}}
\prod_{i\in\mathcal I}
\left[
p_{i,k_i}^{m}
e^{r\sigma B_{j,i,k_i}^{m}(x)}
\right]\\
&=
e^{-r\sigma c_j(x)}
\sum_{m\in\mathcal M}\pi_m
\prod_{i\in\mathcal I}
\left[
\sum_{k=0}^{T+1}
p_{i,k}^{m}
e^{r\sigma B_{j,i,k}^{m}(x)}
\right]
\end{aligned}
\label{eq.12}
\end{equation}
The first equality substitutes the conditional joint probabilities and separates the exponential of the additive feeder contributions. The second replaces the joint sum by a product of feeder-level sums.

\begin{figure}[h]
\centering
\includegraphics[width=3.5in]{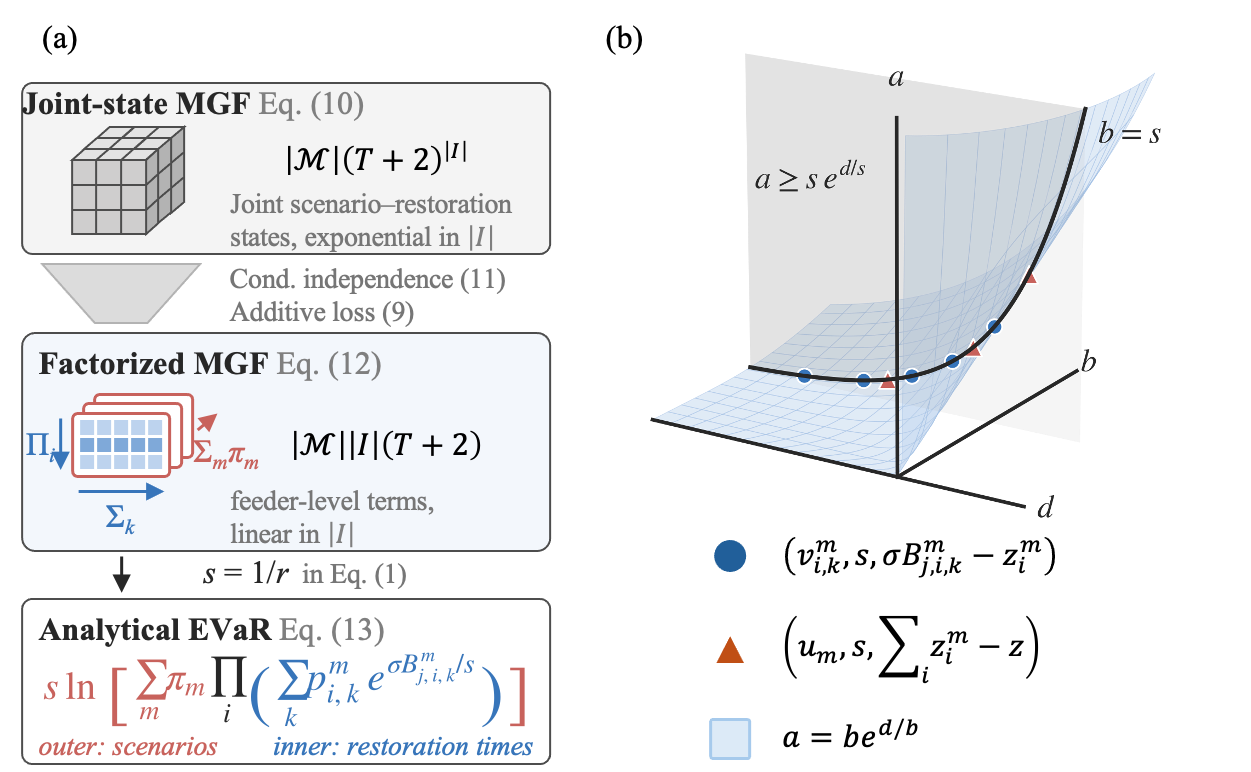}
\caption{Analytical representation and exponential-cone reformulation of the trajectory-based compound risk.}
\label{fig.2}
\end{figure}

Equation~\eqref{eq.12} provides a closed-form analytical expression for the MGF of the compound operating deviation. For given $x$ and $r$, it involves $|\mathcal M||\mathcal I|(T+2)$ feeder-level terms without constructing the joint restoration distribution explicitly. Each feeder's temporal dependence is retained through $B_{j,i,\tau_i}^{m}(x)$, while the outer summation over $m$ preserves dependence across feeders induced by the shared hazard scenario. As illustrated in Fig.~\ref{fig.2}(a), this factorization replaces the joint-state enumeration with $|\mathcal M|$ stacked feeder-level tables, each summed over restoration times and multiplied across feeders, and leads to the nested log-sum-exp form in \eqref{eq.13}.

Substituting the analytical MGF Eq.~\eqref{eq.12} into Eq.~\eqref{eq.1} and setting $s=1/r$ gives:
\begin{equation}
\begin{aligned}
\mathcal R_{\sigma,j}(x)
&:=
\operatorname{EVaR}_{1-\alpha}
\bigl(\sigma Y_j(x,\omega)\bigr)\\
&=
\min_{s>0}
\Bigg\{
-\sigma c_j(x)-s\ln\alpha\\
&\quad+
s\ln\Bigg[
\sum_{m\in\mathcal M}\pi_m
\prod_{i\in\mathcal I}
\left(
\sum_{k=0}^{T+1}
p_{i,k}^{m}
e^{\sigma B_{j,i,k}^{m}(x)/s}
\right)
\Bigg]
\Bigg\}
\end{aligned}
\label{eq.13}
\end{equation}
This formulation is exact under the specified scenario model and conditional independence assumption. For a given scheduling decision $x$, compound risk evaluation reduces to a scalar optimization over $s$, with the MGF evaluated analytically from the feeder-level quantities.

The resulting operating risk is constrained by:
\begin{equation}
\mathcal R_{\sigma,j}(x)
\leq h_{\sigma,j}(x),
\qquad \sigma\in\{-1,+1\}
\label{eq.14}
\end{equation}
where $h_{\sigma,j}(x)$ denotes the available operating margin in direction $\sigma$, assumed affine in $x$.
\subsection{Equivalent Exponential-Cone Reformulation}
\label{sectionIII-C}
The analytical risk expression contains two nested log-sum-exp structures associated with feeder restoration times and hazard scenarios. We reformulate these structures separately to obtain an equivalent conic representation of $\mathcal R_{\sigma,j}(x)\leq h_{\sigma,j}(x)$. The derivation considers a fixed pair $(\sigma,j)$, whose indices are omitted from the auxiliary variables for brevity. Each risk constraint has its own auxiliary variables, with a common scale $s$ across all feeders and scenarios. Zero-probability scenarios are discarded, and $\mathcal K_i^m:=\{k:p_{i,k}^{m}>0\}$ denotes the support of the conditional restoration distribution.

For $s>0$, define the feeder-level log-sum-exp term and the resulting risk objective as:
\begin{subequations}\label{eq.15}
\begin{align}
\ell_i^m(x,s)
&:=
s\ln\left[
\sum_{k\in\mathcal K_i^m}
p_{i,k}^{m}
e^{\sigma B_{j,i,k}^{m}(x)/s}
\right]
\label{eq.15a}\\
F_{\sigma,j}(x,s)
&:=
s\ln\left[
\sum_{m\in\mathcal M}\pi_m
e^{\sum_{i\in\mathcal I}\ell_i^m(x,s)/s}
\right]
\nonumber\\
&\quad-\sigma c_j(x)-s\ln\alpha
\label{eq.15b}
\end{align}
\end{subequations}
The analytical risk expression therefore satisfies $\mathcal R_{\sigma,j}(x)=\inf_{s>0}F_{\sigma,j}(x,s)$.

Introducing an auxiliary variable $z_i^m$ for each feeder and scenario gives the following equivalent transformations:
\begin{equation}
\begin{aligned}
\ell_i^m(x,s)\leq z_i^m
&\iff
\sum_{k\in\mathcal K_i^m}
p_{i,k}^{m}
e^{(\sigma B_{j,i,k}^{m}(x)-z_i^m)/s}
\leq1\\
&\iff
\sum_{k\in\mathcal K_i^m}
p_{i,k}^{m}s
e^{(\sigma B_{j,i,k}^{m}(x)-z_i^m)/s}
\leq s
\end{aligned}
\label{eq.16}
\end{equation}
Define the closed exponential cone as:
\begin{equation}
\begin{aligned}
K_{\exp}
:={}&
\left\{
(a,b,d)\in\mathbb R^3:
b>0,\ a\geq b e^{d/b}
\right\}\\
&\cup
\left\{
(a,0,d)\in\mathbb R^3:
a\geq0,\ d\leq0
\right\}
\end{aligned}
\label{eq.17}
\end{equation}
By introducing $v_{i,k}^{m}$ to bound each scaled exponential term, Eq.~\eqref{eq.16} is equivalent, for $s>0$, to:
\begin{equation}
\begin{aligned}
\left(
v_{i,k}^{m},s,
\sigma B_{j,i,k}^{m}(x)-z_i^m
\right)
&\in K_{\exp},
\quad k\in\mathcal K_i^m\\
\sum_{k\in\mathcal K_i^m}
p_{i,k}^{m}v_{i,k}^{m}
&\leq s
\end{aligned}
\label{eq.18}
\end{equation}
for every $i\in\mathcal I$ and $m\in\mathcal M$.

The outer hazard-scenario mixture is treated similarly. Introducing an auxiliary variable $z$, we obtain:
\begin{equation}
\begin{aligned}
&s\ln\left[
\sum_{m\in\mathcal M}\pi_m
e^{\sum_{i\in\mathcal I}z_i^m/s}
\right]\leq z\\
&\quad\iff
\sum_{m\in\mathcal M}\pi_m s
e^{(\sum_{i\in\mathcal I}z_i^m-z)/s}
\leq s
\end{aligned}
\label{eq.19}
\end{equation}
With auxiliary variables $u_m$, this inequality is equivalently represented by:
\begin{equation}
\begin{aligned}
\left(
u_m,s,\sum_{i\in\mathcal I}z_i^m-z
\right)
&\in K_{\exp},
\quad m\in\mathcal M\\
\sum_{m\in\mathcal M}\pi_m u_m
&\leq s
\end{aligned}
\label{eq.20}
\end{equation}

Combining the two layers and including the zero-scale boundary yields the following conic system:
\begin{subequations}\label{eq.21}
\begin{align}
&\left(
v_{i,k}^{m},s,
\sigma B_{j,i,k}^{m}(x)-z_i^m
\right)\in K_{\exp},
&&\forall m,i,k\in\mathcal K_i^m
\label{eq.21a}\\
&\sum_{k\in\mathcal K_i^m}
p_{i,k}^{m}v_{i,k}^{m}\leq s,
&&\forall m,i
\label{eq.21b}\\
&\left(
u_m,s,\sum_{i\in\mathcal I}z_i^m-z
\right)\in K_{\exp},
&&\forall m
\label{eq.21c}\\
&\sum_{m\in\mathcal M}\pi_m u_m\leq s
\label{eq.21d}\\
&z-\sigma c_j(x)-s\ln\alpha
\leq h_{\sigma,j}(x)
\label{eq.21e}\\
&s\geq0
\label{eq.21f}
\end{align}
\end{subequations}

For any fixed $s>0$, Eq.~\eqref{eq.18} implies $z_i^m\geq\ell_i^m(x,s)$. By the monotonicity of the outer log-sum-exp function, Eq.~\eqref{eq.20} then implies:
\begin{equation}
z\geq
s\ln\left[
\sum_{m\in\mathcal M}\pi_m
e^{\sum_{i\in\mathcal I}\ell_i^m(x,s)/s}
\right]  
\label{eq.22}
\end{equation}
Consequently, the final inequality in Eq.~\eqref{eq.21} ensures $F_{\sigma,j}(x,s)\leq h_{\sigma,j}(x)$. Conversely, whenever this bound holds for $s>0$, a feasible set of auxiliary variables can be constructed as:
\begin{subequations}\label{eq.23}
\begin{align}
z_i^m
&=\ell_i^m(x,s)
\label{eq.23a}\\
v_{i,k}^{m}
&=s e^{(\sigma B_{j,i,k}^{m}(x)-z_i^m)/s}
\label{eq.23b}\\
z
&=s\ln\left[
\sum_{m\in\mathcal M}\pi_m
e^{\sum_{i\in\mathcal I}z_i^m/s}
\right]
\label{eq.23c}\\
u_m
&=s e^{(\sum_{i\in\mathcal I}z_i^m-z)/s}
\label{eq.23d}
\end{align}
\end{subequations}
These choices make both probability-weighted sum constraints hold with equality, establishing equivalence for every positive scale. Geometrically, as shown in Fig.~\ref{fig.2}(b), the inner cones in (21a) and the outer cones in (21c) are slices of the same set $\mathcal K_{\exp}$ at the common scale $b=s$, and under \eqref{eq.23} every cone argument lies on the boundary $a=s\,e^{d/s}$.

The infimum defining EVaR may be attained only as $s\downarrow0$. The closed cone in Eq.~\eqref{eq.17} includes this boundary. Therefore, $\mathcal R_{\sigma,j}(x)\leq h_{\sigma,j}(x)$ holds if and only if auxiliary variables satisfying Eq.~\eqref{eq.21} exist.

Since $B_{j,i,k}^{m}(x)$, $c_j(x)$, and $h_{\sigma,j}(x)$ are affine in $x$, all cone arguments and remaining inequalities in Eq.~\eqref{eq.21} are affine. The reformulated continuous risk constraints are therefore convex. Each risk constraint requires $\sum_{m\in\mathcal M}\sum_{i\in\mathcal I}|\mathcal K_i^m|+|\mathcal M|$ exponential cones, preserving the compact structure of the analytical formulation. These constraints can be incorporated into the UC model to obtain a mixed-integer convex formulation with exponential-cone constraints.

\section{Risk-Aware Electricity Dispatch With Distributed Renewable Integration and Energy Storage}
\label{sectionIV}
\subsection{Distributed PV Generation Decay Model}
\label{sectionIV-A}
TCs affect distributed PV generation through wind-induced panel damage and cloud-induced irradiance reduction. The damage probability $p_i^{\mathrm{PV}}$ is evaluated by a rooftop PV fragility function at the maximum local gust wind $W_i^{\max}$ during the initial TC exposure, and the irradiance reduction is represented by a solar decay factor $\delta_{i,t}^{m}\in[0,1]$ from the attenuation model in~\cite{ceferino2022stochastic}, which relates it to hurricane intensity, distance from the storm center, and storm size. With PV repairs excluded, the mean available PV output of feeder $i$ under scenario $m$ is:
\begin{equation}
P_{i,t}^{\mathrm{PV},m}
=
\delta_{i,t}^{m}
f_{i,t}^{\mathrm{PV}}
\left(1-p_i^{\mathrm{PV}}\right)
C_i^{\mathrm{PV}}
\label{eq.24}
\end{equation}
where $C_i^{\mathrm{PV}}$ is the installed capacity and $f_{i,t}^{\mathrm{PV}}$ is the baseline capacity factor under maximum power point tracking~\cite{subudhi2013comparative}. Eq.~\eqref{eq.24} gives the PV output when the feeder is connected, while its connection status is captured by $\widetilde S_{i,t}$ in Section~\ref{sectionIII-B}. The same approach is applied to utility-scale PV plants with ground-mounted fragility parameters~\cite{ceferino2022stochastic,ceferino2023bayesian}.
\subsection{Compound Risk-Constrained Electricity Dispatch}
\label{SectionIV-B}
\subsubsection{Risk-aware unit commitment model}
Following~\cite{xu2025riskaware}, we deploy a post-event risk-aware UC optimization model to address risk arising from distribution networks with large-scale distributed PV integration under compound climate extremes. The model incorporates net-demand uncertainty jointly driven by feeder restoration and subsequent weather conditions into generator-output, line-flow, intertemporal ramping, and SOC constraints using the analytical risk formulation developed in Section~\ref{sectionIII}. 

The scheduling decisions include generator commitment, startup and shutdown statuses, scheduled generation outputs, storage charging and discharging powers and statuses, stored energy levels, and balancing participation factors of generators and storage units. These decisions are determined after the initial TC exposure, based on the observed feeder states, and before the uncertain restoration times and subsequent weather conditions are realized. For feeder $i$, the connected-state net demand under scenario $m$ is $q_{i,t}^{m}=P_{i,t}^{\mathrm{demand},m}-P_{i,t}^{\mathrm{PV},m}$, where $P_{i,t}^{\mathrm{demand},m}$ accounts for weather-dependent demand and $P_{i,t}^{\mathrm{PV},m}$ is given by Eq.~\eqref{eq.24}. The expected connected net demand $\bar D_{i,t}$ defined in Section~\ref{sectionIII-B} provides the baseline for scheduling.

To accommodate deviations from this baseline, participating generators and storage units adjust their outputs according to an affine balancing policy:
\begin{subequations}
\label{eq.25}
\begin{align}
&\widetilde P_{g,t}(x,\omega)=P_{g,t}+\beta_{g,t}\Delta\widetilde P_t^{\mathrm{net}}(\omega)
\label{eq.25a}\\
&\widetilde P_{s,t}^{\mathrm{dis}}(x,\omega)=P_{s,t}^{\mathrm{dis}}+\beta_{s,t}^{\mathrm{dis}}\Delta\widetilde P_t^{\mathrm{net}}(\omega)
\label{eq.25b}\\
&\widetilde P_{s,t}^{\mathrm{ch}}(x,\omega)=P_{s,t}^{\mathrm{ch}}-\beta_{s,t}^{\mathrm{ch}}\Delta\widetilde P_t^{\mathrm{net}}(\omega)
\label{eq.25c}
\end{align}
\end{subequations}
where $P_{g,t}$ is the scheduled output of generator $g$, $\beta_{g,t}$ is its balancing participation factor, and $\Delta\widetilde P_t^{\mathrm{net}}=\sum_{i\in\mathcal I}\Delta\widetilde P_{i,t}^{\mathrm{net}}$ is the aggregate net-demand deviation. 
The resulting generator-output, line-flow, ramping, and SOC deviations are expressed in the form of $Y_j(x,\omega)$ in Section~\ref{sectionIII-B}. Their directional EVaR constraints limit the risk of the resulting operating deviations exceeding the available operating margins. Energy storage units participate in balancing in a similar manner.
\subsubsection{Optimization formulation}
Let $\mathcal G$, $\mathcal G^{\mathrm R}\subseteq\mathcal G$, $\mathcal S$, $\mathcal L$, and $\mathcal T$ denote the sets of utility-scale generation units, renewable units, energy storage units, power lines, and timesteps, respectively. Distributed PV generation is represented through feeder net demand. The objective function is:
\begin{equation}
\begin{aligned}
\min_{X}\quad
& \sum_{t\in\mathcal T}
\sum_{g\in\mathcal G\setminus\mathcal G^{\mathrm R}}
\Big(
c_gP_{g,t}
+c_g^{\mathrm{fixed}}u_{g,t}
+c_g^{\mathrm{startup}}v_{g,t}
\Big)\\
&+\sum_{t\in\mathcal T}\sum_{s\in\mathcal S}
\Big(
c_s^{\mathrm{dis}}P_{s,t}^{\mathrm{dis}}
+c_s^{\mathrm{ch}}P_{s,t}^{\mathrm{ch}}
\Big)\\
X={}&\{u_{g,t},v_{g,t},w_{g,t},P_{g,t},\beta_{g,t},F_{\ell,t},\\
&\ \ z_{s,t},P_{s,t}^{\mathrm{dis}},P_{s,t}^{\mathrm{ch}},E_{s,t},
\beta_{s,t}^{\mathrm{dis}},\beta_{s,t}^{\mathrm{ch}},\\
&\ \ \forall g\in\mathcal G,\ell\in\mathcal L,s\in\mathcal S,t\in\mathcal T\}
\end{aligned}
\label{eq.26}
\end{equation}
where $X$ collects the binary commitment, startup, and shutdown statuses $u_{g,t}$, $v_{g,t}$, and $w_{g,t}$, scheduled outputs $P_{g,t}$, participation factors $\beta_{g,t}$, and line flows $F_{\ell,t}$; for storage unit $s$, it includes the binary discharging status $z_{s,t}$, discharging and charging powers $P_{s,t}^{\mathrm{dis}}$ and $P_{s,t}^{\mathrm{ch}}$, stored energy $E_{s,t}$, and participation factors $\beta_{s,t}^{\mathrm{dis}}$ and $\beta_{s,t}^{\mathrm{ch}}$. The parameters $c_g$, $c_g^{\mathrm{fixed}}$, and $c_g^{\mathrm{startup}}$ are the marginal (\$/MWh), fixed, and startup costs of generator $g$, and $c_s^{\mathrm{dis}}$ and $c_s^{\mathrm{ch}}$ are the discharging and charging costs (\$/MWh) of storage unit $s$.

The startup and shutdown logic and minimum up- and down-time requirements are given by Eq.~\eqref{eq.27}, where $T_g^{\mathrm U}$ and $T_g^{\mathrm D}$ are the minimum up and down times. Renewable units are always committed.
\begin{subequations}
\label{eq.27}
\begin{align}
&u_{g,t}-u_{g,t-1}=v_{g,t}-w_{g,t}, \forall g\in\mathcal{G}\setminus\mathcal{G}^{\mathrm R},t\in\mathcal{T}
\label{eq.27a}\\
&\sum_{t'=t-T_g^{\mathrm U}+1}^{t}v_{g,t'}\leq u_{g,t}, \forall g\in\mathcal{G}\setminus\mathcal{G}^{\mathrm R},t=T_{g}^{\mathrm U},\dots,T
\label{eq.27b}\\
&\sum_{t'=t-T_g^{\mathrm D}+1}^{t}w_{g,t'}\leq1-u_{g,t}, \forall g\in\mathcal{G}\setminus\mathcal{G}^{\mathrm R},t=T_{g}^{\mathrm D},\dots,T
\label{eq.27c}\\
&u_{g,t},v_{g,t},w_{g,t}\in\{0,1\}, \forall g\in\mathcal{G}\setminus\mathcal{G}^{\mathrm R},t\in\mathcal{T}
\label{eq.27d}\\
&u_{g,t}=1,\quad v_{g,t}=w_{g,t}=0, \forall g\in\mathcal{G}^{\mathrm R},t\in\mathcal{T}
\label{eq.27e}
\end{align}
\end{subequations}

The ramping risk constraints in Eq.~\eqref{eq.28} limit the change in actual generator output between consecutive periods under the balancing policy in Eq.~\eqref{eq.25}, where $R_g^{\mathrm U}$, $R_g^{\mathrm D}$, $P_g^{\max}$, and $P_g^{\min}$ are the ramp-up rate, ramp-down rate, and maximum and minimum capacities, and $\Delta\widetilde P_0^{\mathrm{net}}=0$.
\begin{subequations}
\label{eq.28}
\begin{align}
&P_{g,t}-P_{g,t-1}
+\mathrm{EVaR}_{1-\alpha}\big(\beta_{g,t}\Delta\widetilde P_t^{\mathrm{net}}-\beta_{g,t-1}\Delta\widetilde P_{t-1}^{\mathrm{net}}\big)\nonumber\\
&\quad\le\big(P_g^{\min}+R_g^{\mathrm U}\Delta t\big)u_{g,t}-P_g^{\min}u_{g,t-1}\nonumber\\
&\quad\quad+\big(P_g^{\max}-P_g^{\min}-R_g^{\mathrm U}\Delta t\big)v_{g,t}
\label{eq.28a}\\
&P_{g,t-1}-P_{g,t}
+\mathrm{EVaR}_{1-\alpha}\big(\beta_{g,t-1}\Delta\widetilde P_{t-1}^{\mathrm{net}}-\beta_{g,t}\Delta\widetilde P_t^{\mathrm{net}}\big)\nonumber\\
&\quad\le\big(P_g^{\min}+R_g^{\mathrm D}\Delta t\big)u_{g,t-1}-P_g^{\min}u_{g,t}\nonumber\\
&\quad\quad+\big(P_g^{\max}-P_g^{\min}-R_g^{\mathrm D}\Delta t\big)w_{g,t}
\label{eq.28b}
\end{align}
\end{subequations}

The storage energy dynamics are given by Eq.~\eqref{eq.29}, where $\eta_s^{\mathrm{ch}}$ and $\eta_s^{\mathrm{dis}}$ are the charging and discharging efficiencies and $\Delta t$ is the period length.
\begin{subequations}
\label{eq.29}
\begin{align}
&E_{s,t}=E_{s,t-1}+\big(\eta_s^{\mathrm{ch}}P_{s,t}^{\mathrm{ch}}-P_{s,t}^{\mathrm{dis}}/\eta_s^{\mathrm{dis}}\big)\Delta t
\label{eq.31a}\\
&E_s^{\min}\le E_{s,t}\le E_s^{\max},\quad z_{s,t}\in\{0,1\}
\label{eq.31b}
\end{align}
\end{subequations}

The supply--demand balance and reserve requirement are enforced for the expected connected net demand $\bar D_{i,t}$ in Eq.~\eqref{eq.6}, with reserve factor $R^{\mathrm{res}}$:
\begin{subequations}
\label{eq.30}
\begin{align}
&\textstyle\sum_{g\in\mathcal G}P_{g,t}+\sum_{s\in\mathcal S}\big(P_{s,t}^{\mathrm{dis}}-P_{s,t}^{\mathrm{ch}}\big)=\sum_{i\in\mathcal I}\bar D_{i,t}
\label{eq.30a}\\
&\textstyle\sum_{g\in\mathcal G\setminus\mathcal G^{\mathrm R}}P_g^{\max}u_{g,t}+\sum_{g\in\mathcal G^{\mathrm R}}P_{g,t}+\sum_{s\in\mathcal S}\big(P_{s,t}^{\mathrm{dis}}-P_{s,t}^{\mathrm{ch}}\big)\nonumber\\
&\qquad\ge R^{\mathrm{res}}\textstyle\sum_{i\in\mathcal I}\bar D_{i,t}
\label{eq.30b}
\end{align}
\end{subequations}

The bounds on outputs and participation factors are given by Eq.~\eqref{eq.31}, where $P_s^{\mathrm{dis},\max}$ and $P_s^{\mathrm{ch},\max}$ are the maximum discharging and charging powers. Participation factors are restricted to online generators and to the operating mode of each storage unit, and sum to one so that all realized imbalance is absorbed.
\begin{subequations}
\label{eq.31}
\begin{align}
&P_g^{\min}u_{g,t}\le P_{g,t}\le P_g^{\max}u_{g,t},\quad 0\le\beta_{g,t}\le u_{g,t}
\label{eq.31a}\\
&0\le P_{s,t}^{\mathrm{dis}}\le P_s^{\mathrm{dis},\max}z_{s,t},\quad 0\le P_{s,t}^{\mathrm{ch}}\le P_s^{\mathrm{ch},\max}(1-z_{s,t})
\label{eq.31b}\\
&0\le\beta_{s,t}^{\mathrm{dis}}\le z_{s,t},\quad 0\le\beta_{s,t}^{\mathrm{ch}}\le 1-z_{s,t}
\label{eq.31c}\\
&\textstyle\sum_{g\in\mathcal G}\beta_{g,t}+\sum_{s\in\mathcal S}\big(\beta_{s,t}^{\mathrm{dis}}+\beta_{s,t}^{\mathrm{ch}}\big)=1
\label{eq.31d}
\end{align}
\end{subequations}

The EVaR constraints in Eq.~\eqref{eq.32} limit the risks of actual generator outputs and storage powers violating their bounds after balancing. In response to an upward net-demand deviation, a discharging unit increases its output and a charging unit reduces its consumption.
\begin{subequations}
\label{eq.32}
\begin{align}
&P_{g,t}+\mathrm{EVaR}_{1-\alpha}\big(\beta_{g,t}\Delta\widetilde P_t^{\mathrm{net}}\big)\le P_g^{\max}u_{g,t}
\label{eq.32a}\\
&-P_{g,t}+\mathrm{EVaR}_{1-\alpha}\big(-\beta_{g,t}\Delta\widetilde P_t^{\mathrm{net}}\big)\le -P_g^{\min}u_{g,t}
\label{eq.32b}\\
&P_{s,t}^{\mathrm{dis}}+\mathrm{EVaR}_{1-\alpha}\big(\beta_{s,t}^{\mathrm{dis}}\Delta\widetilde P_t^{\mathrm{net}}\big)\le P_s^{\mathrm{dis},\max}z_{s,t}
\label{eq.32c}\\
&-P_{s,t}^{\mathrm{dis}}+\mathrm{EVaR}_{1-\alpha}\big(-\beta_{s,t}^{\mathrm{dis}}\Delta\widetilde P_t^{\mathrm{net}}\big)\le0
\label{eq.32d}\\
&P_{s,t}^{\mathrm{ch}}+\mathrm{EVaR}_{1-\alpha}\big(-\beta_{s,t}^{\mathrm{ch}}\Delta\widetilde P_t^{\mathrm{net}}\big)\le P_s^{\mathrm{ch},\max}(1-z_{s,t})
\label{eq.32e}\\
&-P_{s,t}^{\mathrm{ch}}+\mathrm{EVaR}_{1-\alpha}\big(\beta_{s,t}^{\mathrm{ch}}\Delta\widetilde P_t^{\mathrm{net}}\big)\le0
\label{eq.32f}
\end{align}
\end{subequations}

Both discharging and charging responses to an upward deviation deplete the stored energy. Unlike the single-period constraints above, the stored energy at time $t$ accumulates the balancing actions of all preceding periods, so the SOC risk depends on the net-demand deviations along the entire preceding trajectory:
\begin{subequations}
\label{eq.33}
\begin{align}
&E_{s,t}+\mathrm{EVaR}_{1-\alpha}\Big(-\Delta t\textstyle\sum_{t'=1}^{t}\gamma_{s,t'}\Delta\widetilde P_{t'}^{\mathrm{net}}\Big)\le E_s^{\max}
\label{eq.33a}\\
&-E_{s,t}+\mathrm{EVaR}_{1-\alpha}\Big(\Delta t\textstyle\sum_{t'=1}^{t}\gamma_{s,t'}\Delta\widetilde P_{t'}^{\mathrm{net}}\Big)\le -E_s^{\min}
\label{eq.33b}
\end{align}
\end{subequations}
where $\gamma_{s,t}=\eta_s^{\mathrm{ch}}\beta_{s,t}^{\mathrm{ch}}+\beta_{s,t}^{\mathrm{dis}}/\eta_s^{\mathrm{dis}}$.

The scheduled line flows and their EVaR constraints are given by Eq.~\eqref{eq.34}, where $H_{\ell-g}$, $H_{\ell-s}$, and $H_{\ell-i}$ are the power transfer distribution factors of line $\ell$ with respect to generator $g$, storage unit $s$, and feeder $i$, and $\kappa_{\ell,i,t}=\sum_{g\in\mathcal G}H_{\ell-g}\beta_{g,t}+\sum_{s\in\mathcal S}H_{\ell-s}(\beta_{s,t}^{\mathrm{dis}}+\beta_{s,t}^{\mathrm{ch}})-H_{\ell-i}$.
\begin{subequations}
\label{eq.34}
\begin{align}
&F_{\ell,t}=\textstyle\sum_{g\in\mathcal G}H_{\ell-g}P_{g,t}+\sum_{s\in\mathcal S}H_{\ell-s}\big(P_{s,t}^{\mathrm{dis}}-P_{s,t}^{\mathrm{ch}}\big)\nonumber\\
&\qquad\ \textstyle-\sum_{i\in\mathcal I}H_{\ell-i}\bar D_{i,t}
\label{eq.34a}\\
&F_{\ell,t}+\mathrm{EVaR}_{1-\alpha}\Big(\textstyle\sum_{i\in\mathcal I}\kappa_{\ell,i,t}\Delta\widetilde P_{i,t}^{\mathrm{net}}\Big)\le F_\ell^{\max}
\label{eq.34b}\\
&{-}F_{\ell,t}+\mathrm{EVaR}_{1-\alpha}\Big(-\textstyle\sum_{i\in\mathcal I}\kappa_{\ell,i,t}\Delta\widetilde P_{i,t}^{\mathrm{net}}\Big)\le -F_\ell^{\min}
\label{eq.34c}
\end{align}
\end{subequations}
Unless otherwise stated, Eqs.~\eqref{eq.28}--\eqref{eq.34} hold for all $g\in\mathcal G$, $s\in\mathcal S$, $\ell\in\mathcal L$, and $t\in\mathcal T$.

All random terms in the EVaR constraints are linear combinations of $\Delta\widetilde P_{i,t}^{\mathrm{net}}$ and thus take the form of $Y_j(x,\omega)$ in Eq.~\eqref{eq.7}. Their EVaR terms are evaluated analytically using Eq.~\eqref{eq.13} and reformulated as exponential-cone constraints through Eq.~\eqref{eq.21}.

\section{Case Study}
\label{sectionV}
\begin{figure}[h]
\centering
\includegraphics[width=\columnwidth]{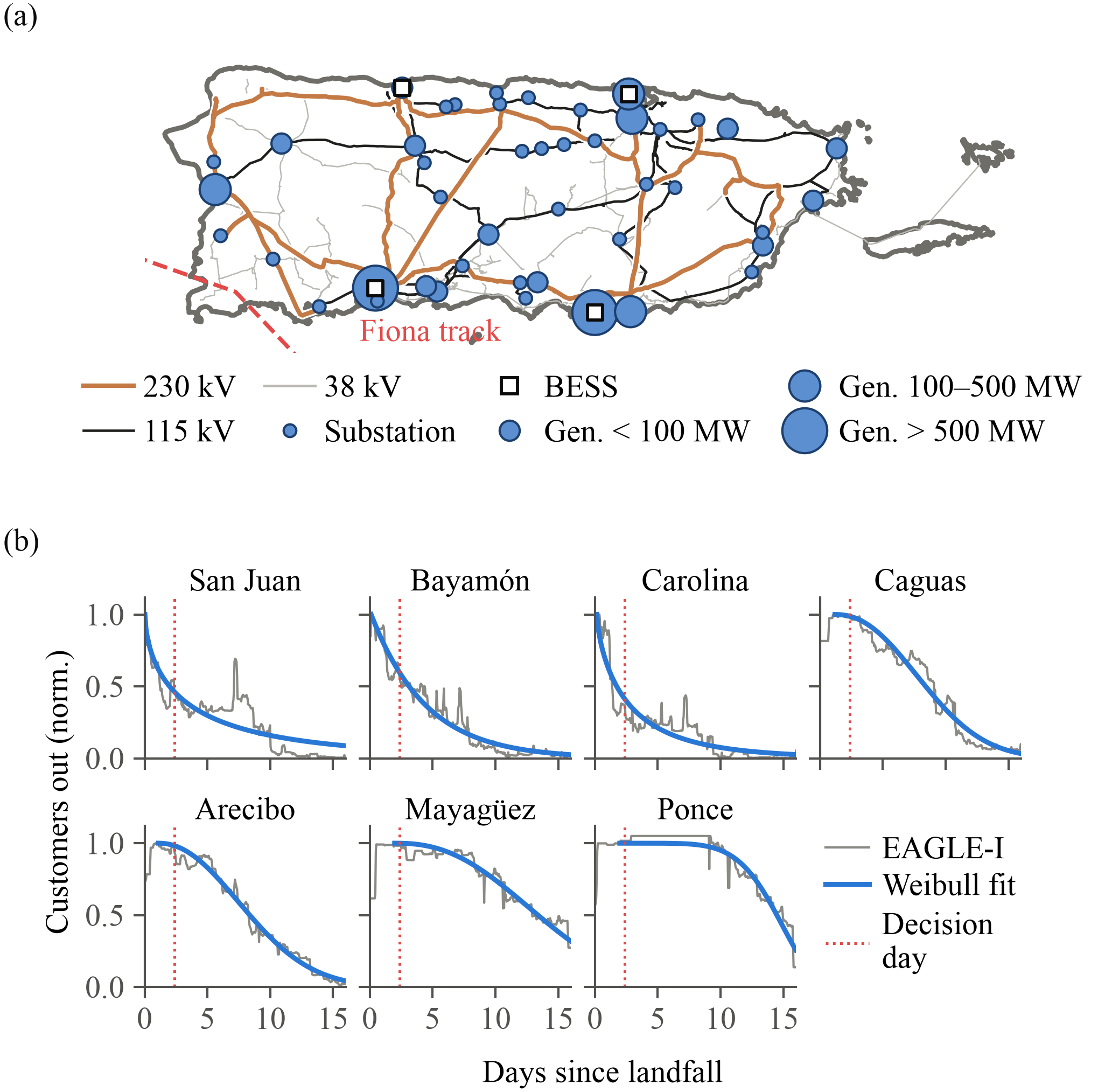}
\caption{Puerto Rico case. (a) Reduced transmission system, generating units, BESS sites, and the track of Hurricane Fiona. (b) Normalized customers out of service in seven regions (EAGLE-I) and the fitted Weibull restoration curves.}
\label{fig.3}
\end{figure}

We consider the case in Puerto Rico on 21~September 2022, 56~h after Hurricane Fiona caused an island-wide blackout~\cite{pasch2023fiona}, when about 0.95~million customers remained without power~\cite{brelsford2024eaglei} and the peak heat index was 3.5\,$^\circ$C above climatology~\cite{hersbach2020era5}.
The reduced 115/230~kV system (Fig.~\ref{fig.3}(a), Table~\ref{tab.I}) has 29 units~\cite{xu2025riskaware}, 430~MW of 4-h battery energy storage system (BESS)~\cite{genera_tesla_bess}, and 948 equivalent feeders in seven regions.
Restoration times are fitted to the regional EAGLE-I outage curves (Fig.~\ref{fig.3}(b)) and slowed by heat-dependent crew productivity~\cite{osha_heat_index}, and $M=5$ hazard scenarios are clustered from ERA5 temperature and humidity of 15--28~September in 1995--2022~\cite{hersbach2020era5}.
\begin{table}[h]
\caption{Puerto Rico Test System}
\label{tab.I}
\centering
\footnotesize
\begin{tabular}{@{}l>{\raggedright\arraybackslash}p{0.74\columnwidth}@{}}
\toprule
Item & Setting \\
\midrule
Network & 102 buses, 128 branches (115/230~kV), 7 monitored lines~\cite{osm} \\
Units & 29 units, 3456~MW~\cite{xu2025riskaware}; online at $t=0$: EcoElectrica, AES~1, San Juan~9~\cite{luma2022sep}; wind and solar not balancing \\
BESS & 120/120/120/70~MW, 4~h~\cite{genera_tesla_bess}; $\eta^{\mathrm{ch}}=\eta^{\mathrm{dis}}=0.92$; SOC 10\%--100\%, initial 50\% \\
Demand & Peak 2751~MW~\cite{xu2025riskaware}; $+2\%/^\circ$C above 27\,$^\circ$C; rooftop PV 600~MW \\
Restoration & Regional Weibull fits to EAGLE-I~\cite{brelsford2024eaglei}; crew productivity 1, 0.85, 0.7, 0.5 by heat-index band~\cite{osha_heat_index} \\
Costs & Balancing \$8/MW-h (units), \$2/MW-h (BESS); flexibility \$189/MWh~\cite{xu2025riskaware} \\
Dispatch & $T=24$~h, $M=5$, $\alpha=0.1$; $10^4$ out-of-sample days \\
\bottomrule
\end{tabular}
\end{table}

Five cases are compared.
Case~1 is the standard UC on $\bar{D}_{i,t}$.
Cases~2--5 solve the EVaR-constrained UC~\eqref{eq.26}--\eqref{eq.34} with $\alpha=0.1$ but represent the compound uncertainty differently:
Case~2 treats feeder states as independent across hours and merges the hazard scenarios, following the hourly risk model of REDUCER~\cite{xu2025riskaware};
Case~3 uses restoration trajectories with merged scenarios;
Case~4 keeps the scenarios with hourly independence;
and Case~5 is the proposed model.
All cases are solved by MOSEK~11.2 through CVXPY~1.9 with a 0.5\% relative mixed-integer programming (MIP) gap on an Apple~M5 computer (10~threads, 24~GB RAM).
Each schedule is evaluated on $10^4$ joint samples of hazard scenario and restoration times drawn from the trajectory model, which is taken as the true distribution.
We report the largest violation frequency among the constraints of each type (excursions above 0.1~MW or MWh), the additional flexibility needed to cover power- and SOC-limit violations (mean and CVaR$_{95}$), and the total cost, i.e., the operating cost plus \$189/MWh for this flexibility~\cite{xu2025riskaware}.

\begin{figure}[h]
\centering
\includegraphics[width=\columnwidth]{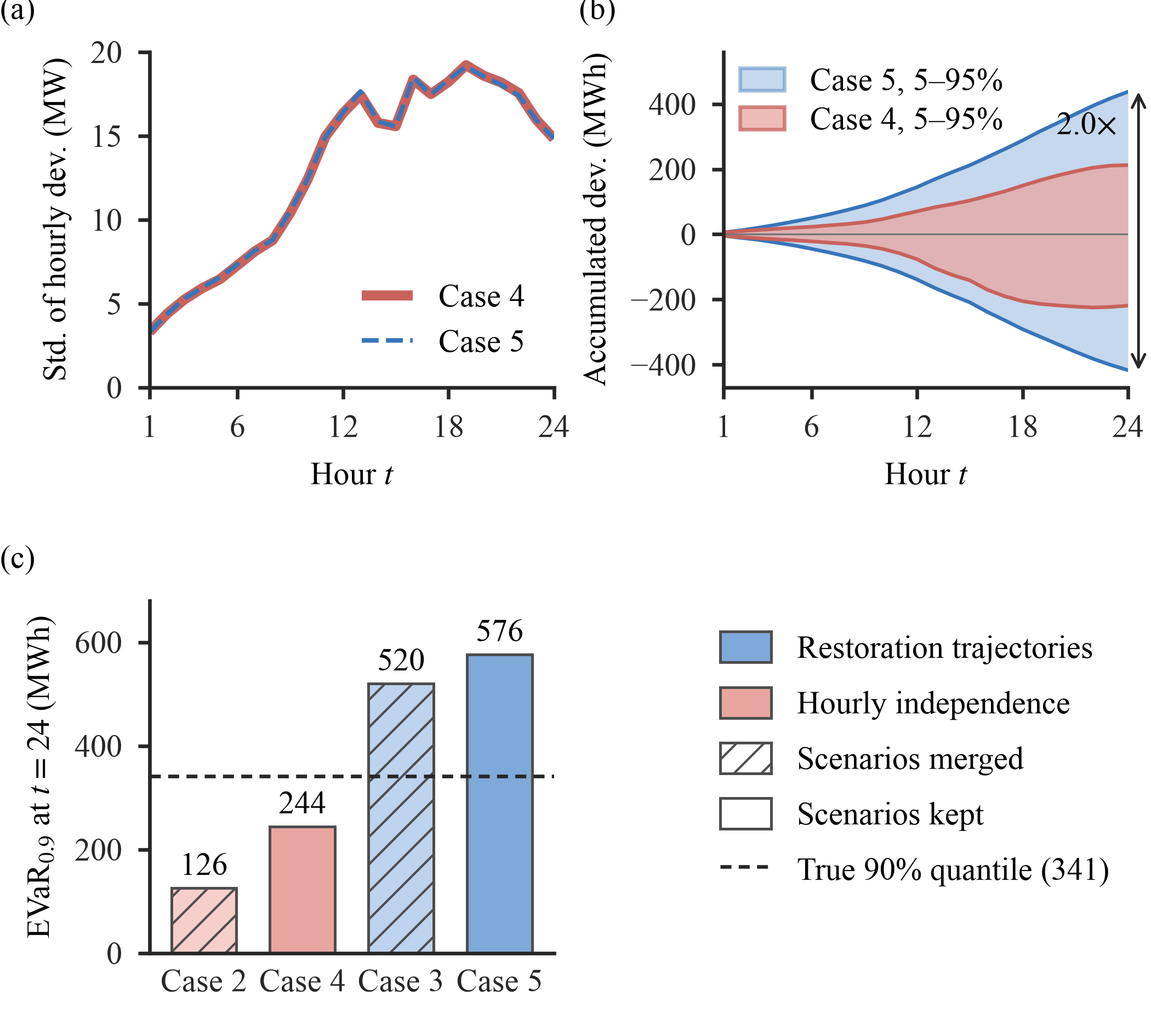}
\caption{System-wide net-demand deviation. (a) Standard deviation of the hourly deviation. (b) 5--95\% range of the accumulated deviation. (c) EVaR$_{0.9}$ of the accumulated deviation at $t=24$ under each case's uncertainty model.}
\label{fig.4}
\end{figure}

Fig.~\ref{fig.4} isolates the effect of restoration trajectories before any dispatch is optimized.
Cases~4 and~5 have identical hourly marginals (Fig.~\ref{fig.4}(a)).
However, a restored feeder stays connected, so the deviations of Case~5 accumulate coherently, and the 5--95\% range of the accumulated deviation at $t=24$ is twice that of Case~4 (Fig.~\ref{fig.4}(b)).
Consequently, the EVaR$_{0.9}$ under hourly independence, 244 and 126~MWh for Cases~4 and~2, falls below the true 90\% quantile of 341~MWh and no longer bounds the risk, whereas the trajectory-based EVaR of Cases~3 and~5, 520 and 576~MWh, remains a valid bound (Fig.~\ref{fig.4}(c)).
This accumulated deviation is what the BESS must absorb.
\begin{figure}[!htbp]
\centering
\includegraphics[width=\columnwidth]{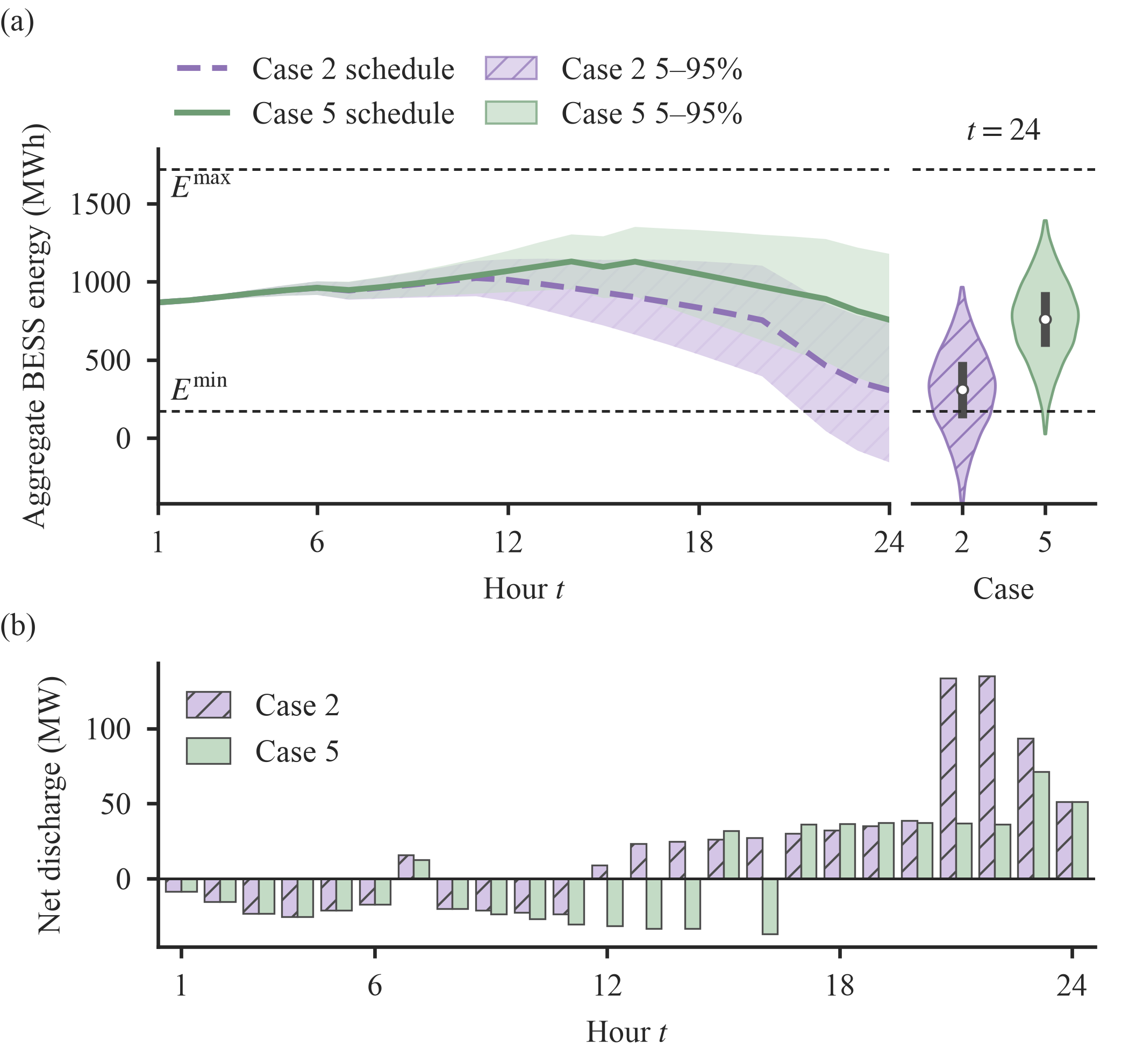}
\caption{BESS dispatch of Cases~2 and~5. (a) Aggregate stored energy: schedule and 5--95\% out-of-sample range, with the out-of-sample distribution at $t=24$ on the right. (b) Scheduled net discharge (positive) and charge (negative).}
\label{fig.5}
\end{figure}
Fig.~\ref{fig.5} shows how Cases~2 and~5 schedule the BESS.
Both cases commit the same units, and in both the BESS carry the entire balancing participation in every hour owing to their lower participation cost (Table~\ref{tab.I}), so the two cases differ only in how the stored energy is positioned.
Case~2 starts discharging at $t=12$ and releases 362~MWh during $t=21$--23, ending the day with 305~MWh, only 133~MWh above $E^{\min}$.
Case~5 keeps charging until $t=14$, raises the stored energy to 1130~MWh, and spreads the discharge over the evening, ending with 756~MWh (Fig.~\ref{fig.5}(b)).
Because Case~5 plans against the accumulated deviation in Fig.~\ref{fig.4}, this energy headroom covers the out-of-sample spread: the aggregate stored energy falls below $E^{\min}$ at $t=24$ in 1.7\% of the samples, against 31\% for Case~2 (Fig.~\ref{fig.5}(a)).

\begin{figure}[!htbp]
\centering
\includegraphics[width=\columnwidth]{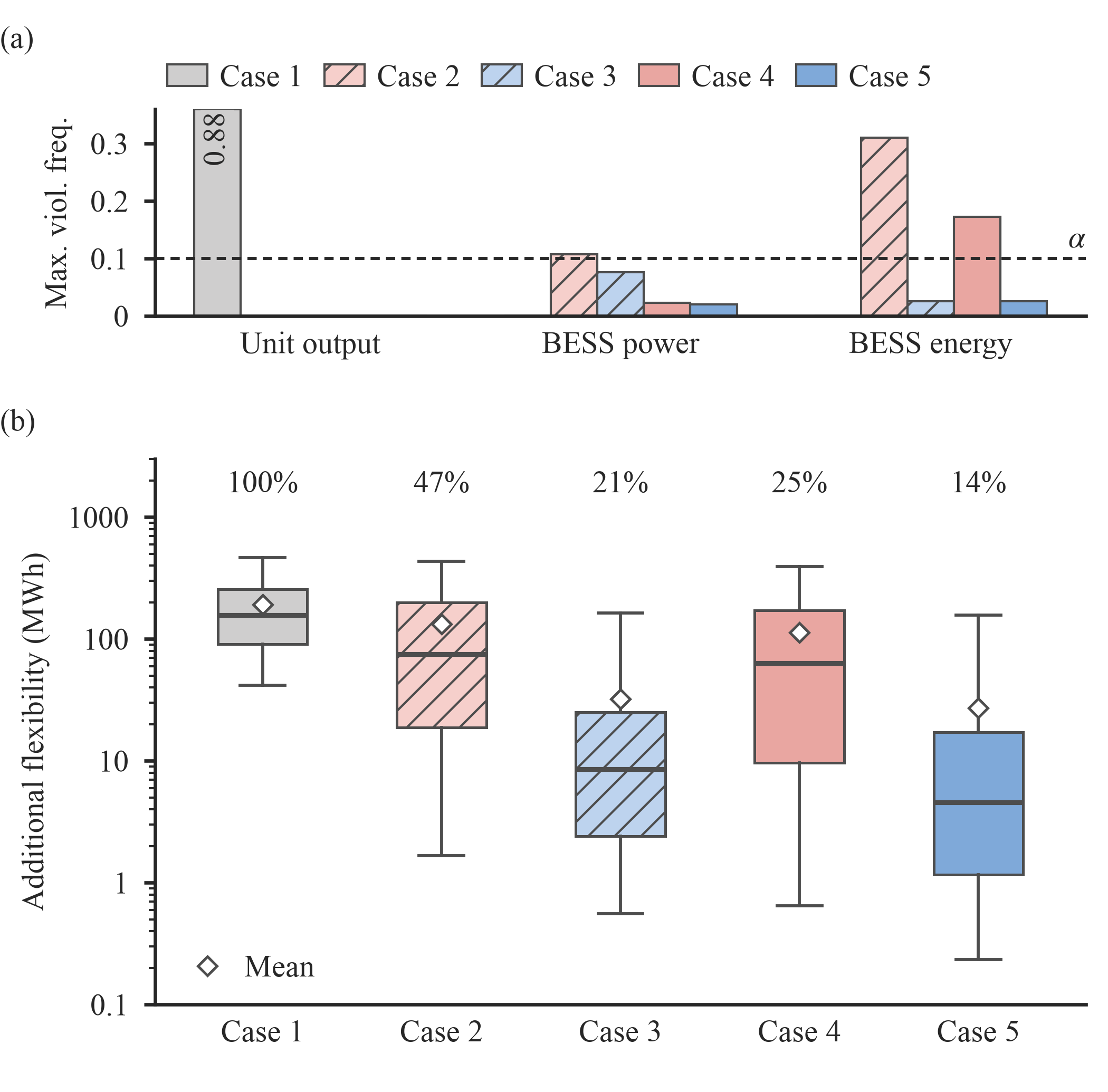}
\caption{Out-of-sample reliability. (a) Largest violation frequency of unit-output, BESS-power, and BESS-energy constraints; ramping and line-flow constraints are never violated. (b) Additional flexibility on the samples that need it (boxes: quartiles; whiskers: 5\% and 95\%; diamonds: means), with the share of such samples on top.}
\label{fig.6}
\end{figure}

Table~\ref{tab.II} and Fig.~\ref{fig.6} extend this comparison to all cases.
Case~1 ignores the uncertainty, so its worst unit-output constraint is violated in 88\% of the samples and every sample needs additional flexibility.
Case~2 protects the unit outputs but not the BESS energy: its worst energy constraint is violated in 31\% of the samples, three times $\alpha$, and keeping the hazard scenarios (Case~4) lowers this only to 17\%.
Only the trajectory-based Cases~3 and~5 keep every constraint within $\alpha$ (Fig.~\ref{fig.6}(a)).
They also shorten the tail: the share of samples needing additional flexibility drops to 21\% and 14\%, and its CVaR$_{95}$ to 110 and 70~MWh, against 469~MWh for Case~2 (Fig.~\ref{fig.6}(b)).
Case~5 has the highest operating cost, 8.7\% above Case~1, but the lowest total cost, 10.8\% below Case~2.
Case~3 reaches a similar total cost, yet its CVaR$_{95}$ is 58\% higher, which is the contribution of the hazard scenarios.
Ramping and line-flow constraints are never violated in this lightly loaded post-event system.
\begin{table}[!t]
\caption{Out-of-Sample Performance ($\alpha=0.1$)}
\label{tab.II}
\centering
\footnotesize
\setlength{\tabcolsep}{2.6pt}
\begin{tabular}{@{}lrrrrrrrr@{}}
\toprule
 & \multicolumn{2}{c}{\begin{tabular}[b]{@{}c@{}}Cost\\(\% of Case~1)\end{tabular}}
 & \multicolumn{2}{c}{\begin{tabular}[b]{@{}c@{}}Flexibility\\(MWh)\end{tabular}}
 & \multicolumn{3}{c}{\begin{tabular}[b]{@{}c@{}}Max.\ violation\\frequency\end{tabular}}
 & \begin{tabular}[b]{@{}c@{}}Time\\(s)\end{tabular} \\
\cmidrule(lr){2-3}\cmidrule(lr){4-5}\cmidrule(lr){6-8}
Case & Oper. & Total & Mean & CVaR$_{95}$ & Unit & BESS $P$ & BESS $E$ & \\
\midrule
1 & 100.0 & 100.0 & 192.5 & 574.8 & \textbf{0.880} & 0 & 0 & 0.3 \\
2 & 105.0 & 78.6 & 61.6 & 468.6 & 0 & \textbf{0.107} & \textbf{0.310} & 236 \\
3 & 107.8 & 70.1 & 6.6 & 109.8 & 0 & 0.076 & 0.026 & 167 \\
4 & 106.5 & 73.4 & 28.4 & 335.6 & 0 & 0.024 & \textbf{0.172} & 1679 \\
5 & 108.7 & 70.1 & 3.7 & 69.5 & 0 & 0.020 & 0.026 & 1207 \\
\bottomrule
\end{tabular}

\vspace{2pt}
\parbox{\columnwidth}{\scriptsize Frequencies are per constraint (unit or BESS, hour); bold: above $\alpha$. Case~1 total cost: \$100.8k per day.}
\end{table}

\begin{figure}[!htbp]
\centering
\includegraphics[width=\columnwidth]{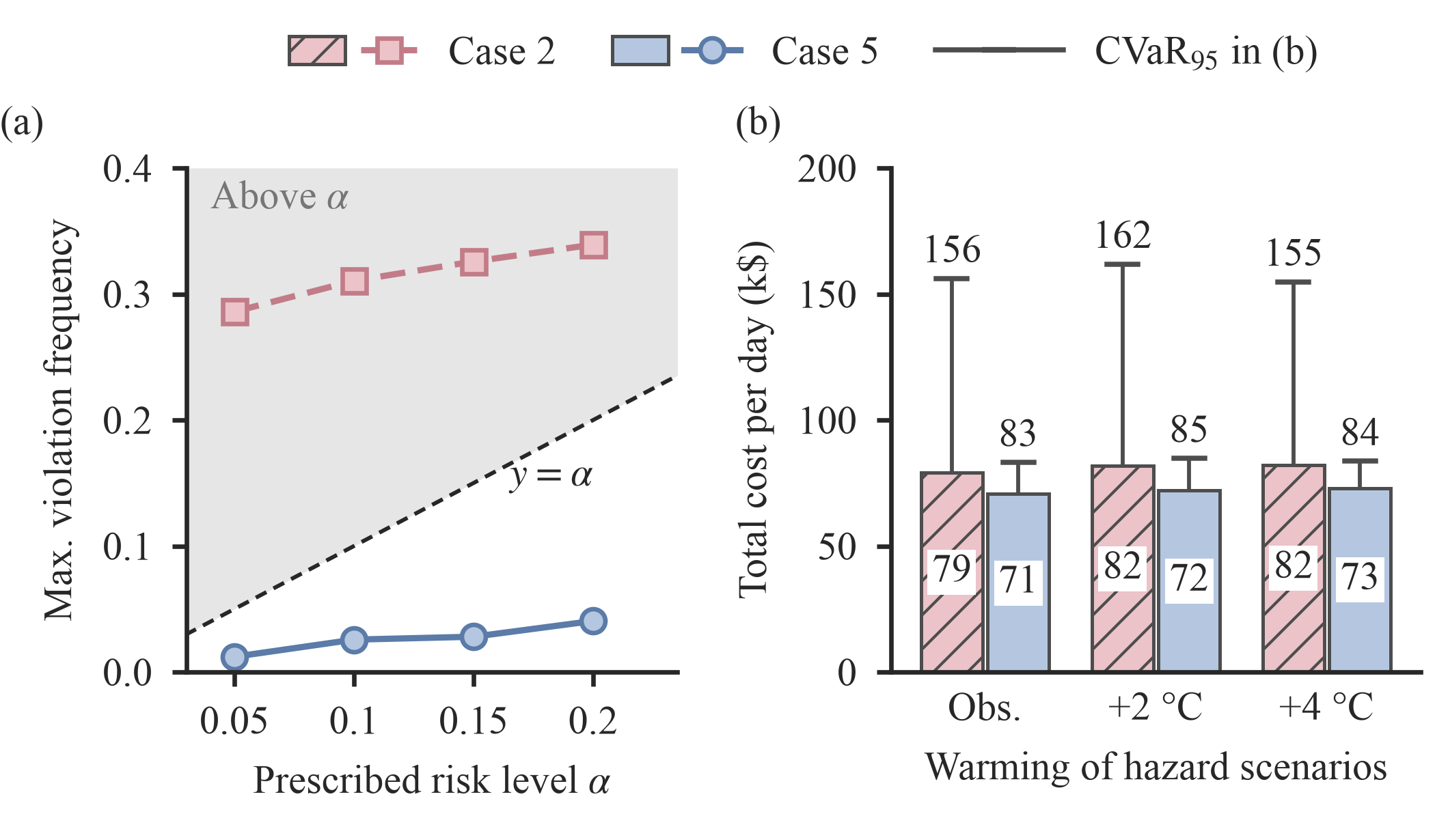}
\caption{Sensitivity of Cases~2 and~5. (a) Largest out-of-sample violation frequency versus the prescribed risk level $\alpha$; the shaded region exceeds $\alpha$. (b) Expected total cost per day (bars) and its CVaR$_{95}$ (caps) with all hazard scenarios warmed by 2 and 4\,$^\circ$C.}
\label{fig.7}
\end{figure}

Fig.~\ref{fig.7} tests whether these conclusions hold under other settings.
As $\alpha$ varies from 0.05 to 0.2, the largest violation frequency of Case~5 remains at 19\%--26\% of $\alpha$, reflecting that EVaR upper-bounds the VaR, whereas that of Case~2 stays at 0.29--0.34 regardless of $\alpha$, because its error stems from the uncertainty model rather than the risk level (Fig.~\ref{fig.7}(a)).
Warming all hazard scenarios by 2 and 4\,$^\circ$C raises the expected total cost of both cases by up to \$2.9k per day, while Case~5 retains an advantage of \$8.6k--9.6k in the mean and \$71k--77k in the CVaR$_{95}$ and keeps every violation frequency below $\alpha$ (Fig.~\ref{fig.7}(b)).
For the warmed scenarios, the MIP of Case~5 stops at the 1-h time limit, so its schedules are feasible but possibly suboptimal.
\begin{figure}[!htbp]
\centering
\includegraphics[width=\columnwidth]{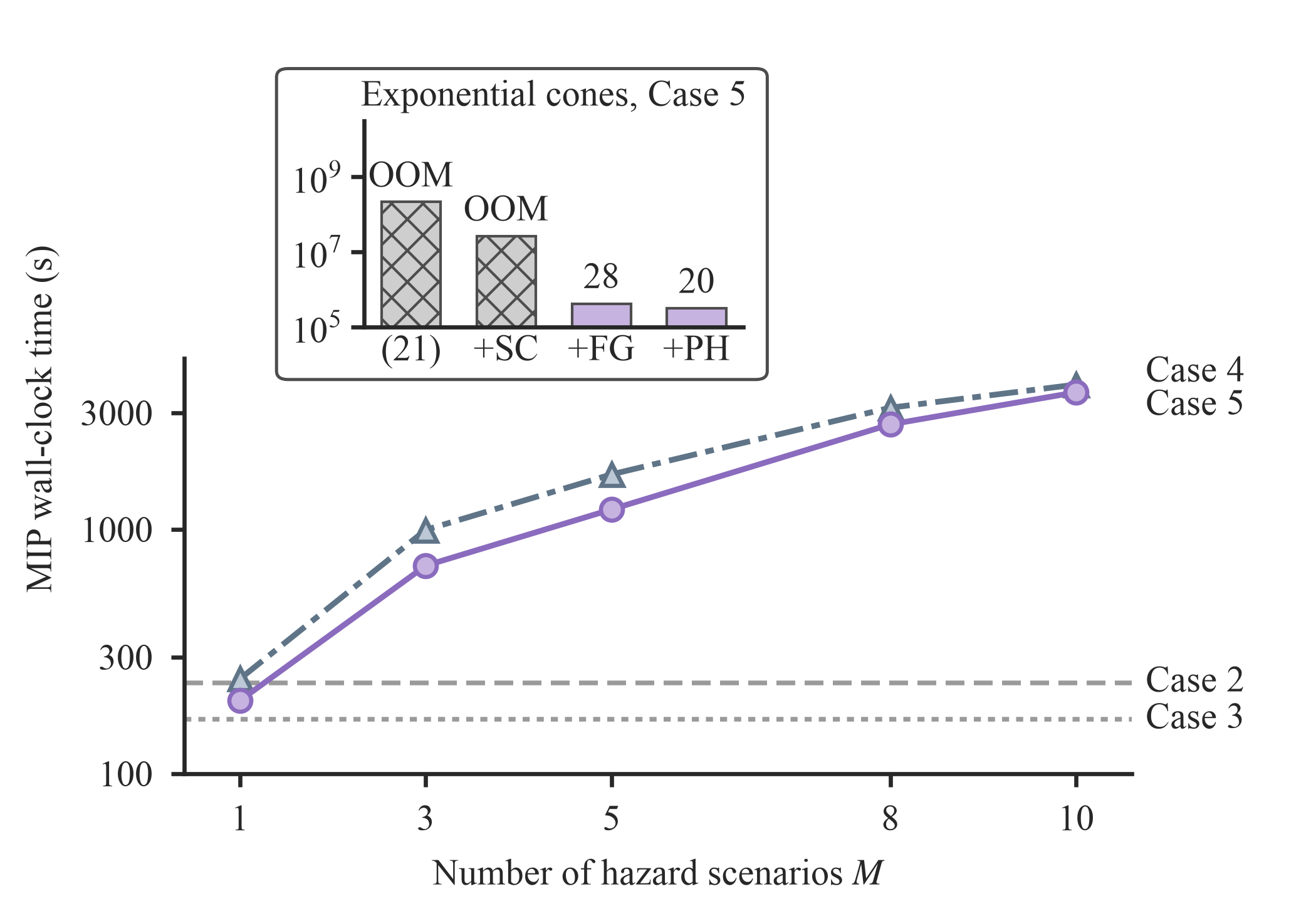}
\caption{MIP wall-clock time (compilation and solution) versus the number of hazard scenarios $M$; Cases~2 and~3 merge the scenarios and are independent of $M$. Inset: exponential cones of Case~5 ($M=5$) under successive exact reductions (SC: support collapse; FG: feeder grouping; PH: positive homogeneity), labeled with MIP time (min); OOM: out of memory (24~GB).}
\label{fig.8}
\end{figure}
Finally, Fig.~\ref{fig.8} examines the computational cost of the proposed formulation.
Written directly as Eq.~\eqref{eq.21} for every feeder and restoration time, the risk constraints of Case~5 require $2.1\times10^{8}$ exponential cones and cannot be built in 24~GB of memory.
Three exact reductions make the model tractable: merging restoration times that yield the same deviation (support collapse) reduces the count to $2.6\times10^{7}$; sharing one set of cones among identical feeders (feeder grouping) reduces it to $4.2\times10^{5}$; and linearizing the single-factor constraints by the positive homogeneity of EVaR reduces it to $3.1\times10^{5}$, cutting the MIP time from 28 to 20~min.
The continuous relaxations before and after the last step have the same optimum, confirming that the reductions are exact.
With the scenarios kept, the solution time grows roughly in proportion to $M$ and reaches about 1~h at $M=10$, which is acceptable for day-ahead post-event scheduling.
Case~5 is faster than Case~4 for every $M$, as the trajectory model needs 23\% fewer cones than hourly independence.

\section{Conclusion}
\label{sectionVI}
This paper has presented a compound climate risk-aware electricity dispatch framework for power systems facing prolonged outages under compound climate extremes.
Unlike hourly risk models, which treat feeder states as independent across hours, the proposed trajectory-based model describes each feeder's connection trajectory through its uncertain restoration time and couples it with subsequent hazard scenarios.
Analytical moment-generating functions and exponential-cone reformulations embed the resulting risk constraints into UC without enumerating joint feeder states, and exact reductions keep the model tractable for a real-world system.
Applied to Puerto Rico after Hurricane Fiona in 2022, the framework shows that hourly models, although matching the hourly statistics, underestimate the risk accumulated in energy storage and violate storage limits in up to 31\% of the samples at a 10\% risk level.
In contrast, the proposed model keeps all violations within this level, reduces the expected total cost by 10.8\%, and markedly shortens the tail of the additional flexibility requirement.
These advantages hold across risk levels and under intensified subsequent hazards, indicating that the temporal structure of restoration should be considered in post-event dispatch.
The framework thus offers system operators a practical tool for scheduling generation and storage during prolonged outages, supported by outage monitoring and weather forecasts.

\section*{Acknowledgments}
This work was assisted by Claude (developed by Anthropic) for initial drafting and language refinement. All content, including methodology, results, and conclusions, was thoroughly reviewed, validated, and revised by the authors to ensure scientific accuracy, originality, and integrity. The use of generative AI tools complies with IEEE’s authorship and transparency guidelines.

\bibliographystyle{IEEEtran}
\bibliography{bib}


 




\vfill

\end{document}